\documentclass[twocolumn, amsmath,superscriptaddress, nohypertex, amsfonts,prb]{revtex4}
\usepackage{graphicx}
\usepackage{url}
\pdfoutput=1
\begin{document}

\title{Percolation with plasticity for neuromorphic systems}

\author{V. G. Karpov}
\email{victor.karpov@utoledo.edu}
\affiliation{Department of Physics and Astronomy, University of Toledo, Toledo, OH 43606, USA}
\author{G. Serpen}\email{gursel.serpen@utoledo.edu}\affiliation{Department of Electrical Engineering and Computer Science, University of Toledo, Toledo, OH 43606, USA}
\author{Maria Patmiou}
\email{maria.patmiou@rockets.utoledo.edu}
\affiliation{Department of Physics and Astronomy, University of Toledo, Toledo, OH 43606, USA}

\date{\today}

\begin{abstract}

We develop a theory of percolation with plasticity media (PWPs) rendering properties of interest for neuromorphic computing. Unlike the standard percolation, they have multiple ($N\gg 1$) interfaces and exponentially large number ($N!$) of conductive pathways between them. These pathways consist of non-ohmic random resistors that can undergo bias induced nonvolatile modifications (plasticity). The neuromorphic properties of PWPs include: multi-valued memory, high dimensionality and nonlinearity capable of transforming input data into spatiotemporal patterns, tunably fading memory ensuring outputs that depend more on recent inputs, and no need for  massive interconnects. A few conceptual examples of functionality here are random number generation, matrix-vector multiplication, and associative memory. Understanding PWP topology, statistics, and operations opens a field of its own calling upon further theoretical and experimental insights.

\end{abstract}
\maketitle
\section{Introduction}\label{sec:intro}

Devices for neuromorphic computing remain among the most active areas of research with a variety of models for neurons, synapses and their networks. \cite{schuman2017,zidan2018,bohnstingal2019,sebastian2019,wan2019,zhang2019,upadhhyay2019} They are typically built of nonvolatile memory cells and interconnects  wired in a certain architecture.

Here we introduce a concept of neuromorphic devices where neither artificial memory cells nor interconnects are required. They are based on disordered  materials with percolation conduction  \cite{efros,shik,snarskii} such as amorphous, polycrystalline, and doped semiconductors, or granular compounds. We recall that percolation transport takes place in systems of microscopic random resistors, and is dominated by the infinite cluster of smallest resistors allowing connectivity between the electrodes.

Among possible percolation conduction materials, we consider those exhibiting plasticity, i. e. exhibiting nonvolatile changes in their resistances in response to strong enough electric field.  They include metal oxides and chalcogenide compounds used with resistive random access memory (RRAM) \cite{lanza2014} and phase change memory (PCM), \cite{sebastian2019} granular metals, \cite{gladskikh2014} and nano-composites.\cite{song2016}

\begin{figure}[t]
\includegraphics[width=0.35\textwidth]{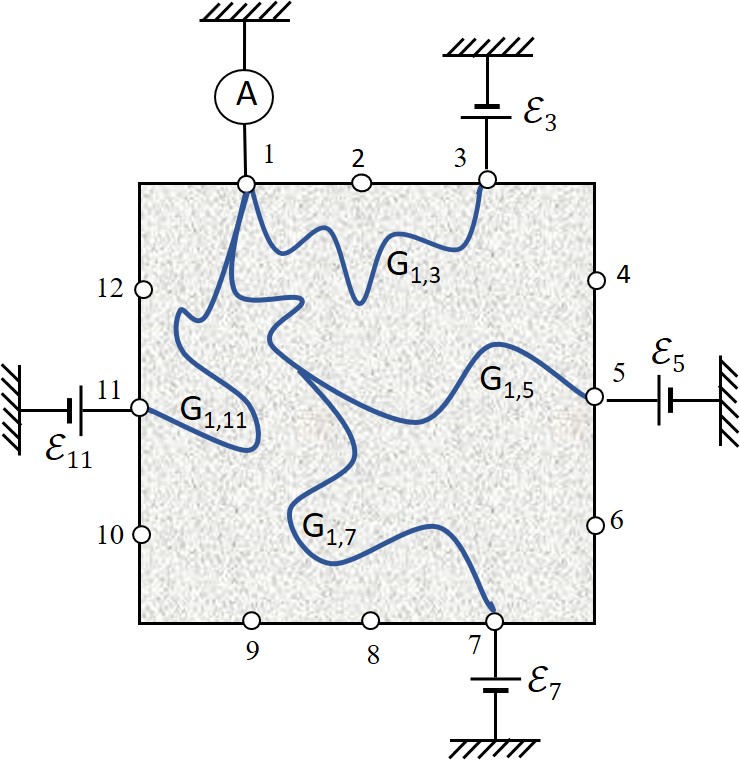}
\caption{Schematic 2D illustration of percolation systems with multiple local interfaces (electrodes). $G_{ij}$ stand for pathway conductances. Only a small number of $12!\approx 10^{13}$ pathways possible in the sketch are shown. The peripheral numbers 1,2,..,12 refer to the assumed 12 electrodes.\label{Fig:PWPconcept}}
\end{figure}

A PWP example in Fig. \ref{Fig:PWPconcept} shows some conductive pathways for the case of a relatively small number of electrodes. Anticipating a particular application below, Fig. \ref{Fig:PWPconcept} assumes certain voltages ${\cal E}_i$ applied to all electrodes but one used to measure the electric current $I$. Other implementations would assume different circuitries with various power sources and meters attached to their multiple electrodes. As explained in what follows, each of the pathways can undergo multiple field induced changes thereby presenting a multivalued memory unit.

We note the following PWP features relevant for neuromorphic applications:\\
(i) The exponentially large combinatorial number, $M=N!\approx \exp(N\ln N)\gg 1$, of interelectrode resistances $R_{ij}$ that scales exponentially with the number ($N\gg 1$) of electrodes. For example, $M\sim 10^{13}$ in a design of Fig. \ref{Fig:PWPconcept}. Such extremely high dimensionality in combination with resistors' nonlinearity (non-ohmicity) makes PWP ideal objects for the reservoir computing. \cite{tanaka2019,seoane2019} \\
(ii) Multivalued memory in conductive pathways operated by electric pulses that modify $R_{ij}$ due to material plasticity; \cite{patmiou2019,karpov2020} they play the role of multiple microscopic memory cells. \\
(iii) Direct connectivity between the bond-forming microscopic resistors eliminates the need for artificial interconnects. In fact, each microscopic resistor in PWP can act as a nanometer sized memristor without artificial interconnects. \\
(iv) The multivalued memory in combination with multiplicity ($M\gg 1$) offers a platform for the in-memory computing. \cite{verma2019} Furthermore, mathematically, series of cells in PWP present multidimensional random vectors forming a base for hyperdimensional computing. \cite{kanerva2009,mitrokhin2019,karunaratne2019}\\
(v) The randomness of PWP topology offers a natural implementation of the randomly wired neural networks outperforming ( at least in some applications) their regularly wired counterparts. \cite{hie2019,zopf2018} That randomness can as well become beneficial with reservoir computing applications. \cite{tanaka2019}

In what follows we consider the physical parameters of PWPs and some examples of their neuromorphic functionality. The paper is organized as follows. It starts with a purely
qualitative discussion in Sec. \ref{sec:qual} that explains without any
math the model and the logic of the paper. Section \ref{sec:stanper}
describes the standard percolation concept. The PWP systems, including their limiting cases of large and small structures, are described in Sec. \ref{sec:PWPs}. Examples of PWP's neuromorphic functionality are presented in Sec. \ref{sec:examp}. Following the nomenclature established for other neuromorphic systems, Sec. \ref{sec:met} briefly discusses certain metrics of the proposed PWP devices. Similarities and architectural differences between PWPs and biological neural networks are discussed in Sec. \ref{sec:bio}. Numerical estimates in Sec. \ref{sec:num} suggest
that the proposed systems can be experimentally implemented allowing verifications of their expected properties. We briefly touch upon the issue of inherent randomness of PWP systems in Sec. \ref{sec:disorder}. The conclusions in Sec. \ref{sec:concl} list this
approach’s capabilities and limitations.

\section{Qualitative description}\label{sec:qual}
This section provides a simplified “low resolution” guide
for subsequent consideration. It offers a brief
summary of our work aimed at the backgrounds of electrical engineering and computer science researchers most significantly contributing to the fields of artificial intelligence and neuromorphic computing.
\begin{enumerate}
\item Underlying our approach is the classical concept of percolation conduction as the electric transport through a network of exponentially different random resistors. It is dominated by the percolation cluster formed by conductive bonds connecting the electrodes as illustrated in Fig. \ref{Fig:perc}. The bonds are composed of the elements with minimum resistances whose total concentration is just sufficient to form the electrode connecting pathways. The characteristic mesh size of the percolation cluster $L_c$ is much greater than the linear size $a$ of a microscopic resistor (see Fig. \ref{Fig:perc}). The percolation cluster is effectively uniform over distances significantly exceeding $L_c$.
 \begin{figure}[t]
\includegraphics[width=0.35\textwidth]{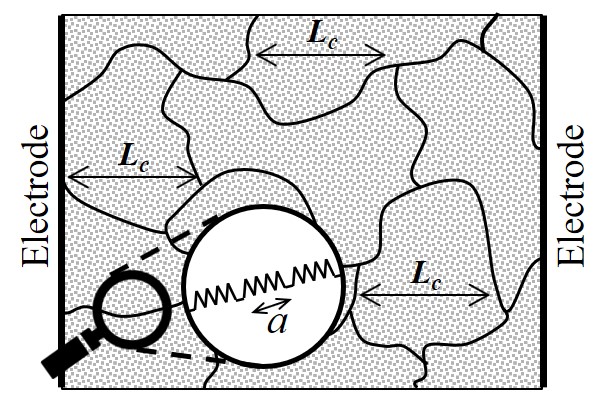}
\caption{A sketch of the percolation cluster between two electrodes. Shown under magnification is a bond domain consisting of microscopic resistances in series. \label{Fig:perc}}
\end{figure}

\item The percolation conduction is extremely nonohmic starting from voltages $U\sim kT/q\sim 0.025-0.05$ V. Such nonohmicity is due to the fact that electric current must be continuous along a percolation bond. Because the latter is composed of multiple ($\sim L_c/a\gg 1$) exponentially different resistors, voltages on some of them must be exponentially high to maintain the same current. As a result, the exponentially higher voltage, of the order of that across the entire $L_c$ distance, is localized on the most resistive element.
\item The local high fields in the percolation cluster can be strong enough to cause local structural transformations (nonvolatile switching) and corresponding decrease in local resistivity. The possibility of such transformations is customarily referred to as plasticity in the neuromorphic domain; hence, percolation with plasticity (PWP). Each drop in resistance caused by plasticity can serve as a memory record detectable relative to the preceding value. Under the same conditions, the voltage originally localized on the highest of the series of exponentially random resistors will concentrate on the next highest one when the former is switched to the low resistance state, etc. causing a sequence of resistance drops.
\item In addition to being a non-ohmic resistance, each element of a percolation bond possesses capacitive properties. It will perform as a capacitor if the electric field pulse is fast enough to make its displacement current greater than the real one. Such capacitive elements can be interpreted as `slow' resistors. These elements possess Maxwell's relaxation times greater than the pulse time. They are intuitively referred to as capacitors because the $RC$ time of an element with resistivity $\rho$, dielectric permittivity $\varepsilon$, length $l$ and area $A$ equals $\varepsilon\rho/4\pi$ thus representing the Maxwell relaxation time. In spite of their large resistances, such capacitive elements do not accommodate significant voltages because their currents are due to the voltage rates of change rather than voltages themselves. These displacement currents are physically related to the local charging processes, the duration of which is reciprocal in the corresponding element's resistance.
\item If pulses arriving at the same element within its relaxation time are in integral strong enough to turn it from the capacitive to resistive mode, the resistance of bond, in which it belongs, will change. The change becomes nonvolatile if the resulting field is sufficiently strong for switching. That property is similar to that of the spike timing dependent plasticity (STDP) \cite{zopf2018} central to neuromorphics.
\item Unlike the standard two-electrode percolation systems, PWPs have multiple ($N\gg 1$) electrodes as illustrated in Fig. \ref{Fig:PWPconcept}. That creates $N!\gg 1$ permutation related interelectrode percolation bonds that can be utilized for memory and computing.
\item In the pulse regime, each bond can accommodate a number (numerically estimated as $\sim 10$) of nonvolatile changes in resistance.
\item Expected PWP applications include multi-valued memory, random number generation, associative learning, and reservoir computing. The parameters of proposed systems fall in the domain of practically implementable material systems.
\end{enumerate}

\section{Standard percolation conduction}\label{sec:stanper}
We start with a recap of the  pertinent percolation concepts. \cite{efros,shik,snarskii} Percolation conduction is dominated by the sparse infinite cluster of the exponentially different random resistors between the two large electrodes. The cluster bonds consist of the minimally strong random resistors with total concentration sufficient to form an infinitely connected network. It is effectively uniform over large distances $L\gg L_c$ where $L_c$ is the correlation radius determining its characteristic mesh size.

Each bond of the cluster consists of a large number of microscopic resistors $R_i$. Their exponential randomness is described as $R_i=R_0\exp(\xi _i)$ where quantities $\xi _i$ are uniformly distributed in the interval $(0,\xi _{\rm max} )$ and $i=1,2,..$.

The physical meaning of $\xi$ depends on the type of a system. For definiteness, we assume here $\xi _i=V_i/kT$ corresponding  to random barriers $V_i$ in noncrystalline materials where $k$ is the Boltzmann's constant and $T$ is the temperature. In reality, the nature of percolation conduction can be more complex including e. g. finite size effects and thermally assisted tunneling between the microscopic sites in nanocomposites. \cite{lin2013,eletskii2015} These complications will not qualitatively change our consideration below.

The cluster constituting microscopic resistors exhibit non-ohmicity due to the field induced suppression of their barriers $V_i$, according to
\begin{equation}\label{eq:nonohm}J_i=J_0\exp(-V_i/kT)\sinh(qU_i/2kT)\end{equation}
where $J_i$ is the resistor current, $J_0$ is a constant, $U_i=E_ia$ is the voltage applied to the barrier, $a$ and $E_i$ are, respectively, the barrier width and local electric field, and $q$ is the electron charge.
\begin{figure}[t!]
\includegraphics[width=0.35\textwidth]{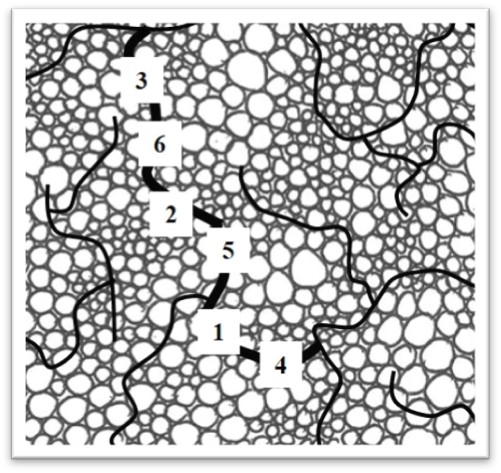}
\caption{A fragment of conductive pathways in the infinite percolation cluster representative of polycrystalline or granular materials. Numbers 1-6 represent random resistors in descending order. \label{Fig:PWP2}}
\end{figure}

An important conceptual point \cite{shklovskii1976,shklovskii1979,aladashvili1989,patmiou2019} is that the applied voltage concentrates first on the strongest resistor of a percolation cluster bond (resistor 1 in Fig. \ref{Fig:PWP2}) suppressing it to the level of the next strongest (resistor 2 in Fig. \ref{Fig:PWP2}), so the two equally dominate the entire bond voltage drop. It then suppresses the next-next strongest resistors (3,4,5,.. in Fig. \ref{Fig:PWP2}), etc. As a result, the percolation cluster changes its structure as
\begin{equation}\label{eq:corrad}L_c=a\sqrt{V_0/qEa},\quad {\rm and}\quad   \Delta V=\sqrt{V_mqaE}\end{equation}
resulting in the macroscopic non-ohmic conductivity,
\begin{equation}\label{eq:shkl}\sigma =\sigma _0\exp\left(\sqrt{\xi_{\rm max}qaE/kT}\right).\end{equation}
Here $L_c$ and $\Delta V$ are, respectively, the field dependent correlation radius and maximum barrier decrease in the percolation cluster, and $V_0$ is the amplitude of barrier variations.

Two assumptions behind the above discussed nonohmicity are: (a) The volatility of bias induced changes where each microscopic resistor adiabatically adjusts its resistance to the instantaneous bias. (b) The quasistatic nature of biasing implying time intervals exceeding the relaxation times of all microscopic resistors, i. e. $t\gg \tau _m=\tau _0\exp(\xi _m)$ where $\tau _0\sim 1$ ps is on the order of the characteristic atomic vibration times.

The quasistatic bias limitation can be relaxed to describe non-ohmicity in the pulse regime. \cite{karpov2020} It was argued that for a cluster bond a bias pulse of length $t$ generates the current
\begin{equation}\label{eq:nonohm}I=I_0\frac{\tau _0}{t}\exp\left(\sqrt{\frac{2\xi _{\rm max}}{N}\frac{qV}{kT}}\right).\end{equation}
This result applies when $t$ is shorter than the maximum relaxation time $\tau _{\rm max}=\tau _0\exp(\xi _{\rm max})$ and is formally different from that of dc analysis \cite{shklovskii1976} by the substitution $\xi _{\rm max}\rightarrow \xi _t\equiv\ln(t/\tau _0)$; the two results coincide when $\xi _t= \xi _{\rm max}$.
For the entire percolation cluster, the  modification $\xi _{\rm max}\rightarrow \xi _t$  predicts the current,
\begin{equation}\label{eq:PFmod}
I=I_0\frac{\tau _0}{t}\exp\left(\sqrt{\frac{2a{\cal E}q}{3kT}\ln \frac{t}{\tau _0}}\right).\end{equation}

To avoid any misunderstanding, we note that the field induced changes in resistances of percolation clusters described in this section is volatile (i. e. it disappears when the field is removed). It should not be confused with the nonvolatile plasticity introduced here.

\section{Percolation with plasticity systems}\label{sec:PWPs}

PWP phenomenon differs from the standard percolation conduction in both topology and non-ohmicity. The former is such that the proposed PWP has multiple ($N\gg 1$) electrodes; the latter is due to the nonvolatile nature of bias induced changes. These properties can be somewhat different for PWP systems with large ($L\gg L_c$) and small ($L\ll L_c$) geometrical dimensions $L$ as described next.

\subsection{General}\label{sec:PWPST}
We start with noting some general properties of PWPs relevant to neuromorphic applications.

(1) The variations $\Delta R_{ij}$ (with respect to the averages $\langle R_{ij}\rangle$) in the interelectrode resistances $R_{ij}$ are random quantities that are uncorrelated with any desired accuracy for not-too-close electrodes. Consider  $\Delta R_{ij}=R_{ij}-\langle R_{ij}\rangle$ with $R_{ij}= R_0\sum _1^{N_{ij}}\exp(\xi _i)$ for a bond of $N_{ij}$ resistors and $\xi _i$ uniformly distributed in the interval $(\xi _{\rm min},\xi _{\rm max})$. It is then straightforward to obtain the correlation coefficient between the resistances of $(i,j)$ and $(k,l)$ bonds,
\begin{equation}\label{eq:corco}
C\equiv \frac{\langle\Delta R_{ij}\Delta R_{kl}\rangle}{\sqrt{\langle(\Delta R_{ij})^2\rangle\langle(\Delta R_{kl})^2\rangle}}=\frac{N_s}{\sqrt{N_{ij}N_{kl}}}
\end{equation}
where $N_{ij}$ and $N_{kl}$ represent the numbers of microscopic resistances in those bonds, and $N_s$ is the number of resistances shared between them. We describe each bond as a random walk. Then, if the two bonds are not close geometrically, separated by distances $L_{ijkl}$ exceeding $a\sqrt{N_{ij}+N_{kl} }$, then their overlap is exponentially small, and
\begin{equation}\label{eq:corco1}C\sim \exp\{-L_{ijkl}^2/[a^2(N_{ij}+N_{kl})]\}\ll 1.\end{equation}
The averages implied by the definition for $\Delta R_{ij}$ can be readily measured for an ensemble of geometrically similar pairs, such as (1,9), (2,8), (3,7), (4,12), etc. in Fig. \ref{Fig:PWPconcept}.

(2) Related to the above item (1), there is a characteristic distance $\overline{l_{ij}}$, above which the electrodes $i$ and $j$ are electrically independent (no crosstalk between them). To estimate that length, we use it in place of  $L_{ijkl}$ in Eq. (\ref{eq:corco1}) setting also $N_{ij}\sim N_{kl}\sim L/a$ where $L$ is the sample size and $a$ is the microscopic resistor length. This yields,
\begin{equation}\label{interel}\overline{l_{ij}}\sim \sqrt{La}. \end{equation}
Assuming as a rough estimate $a\sim 10$ nm and $L\sim 1$ cm, yields $\overline{l_{ij}}\sim 10 $ $\mu$m.

(3) High frequency inductive coupling of conductive pathways can be estimated based on a model of a wire and a loop of diameter $L$. Using the standard electrodynamics, the ratio of the induced current over the primary current then becomes (in the Gaussian system)
\begin{equation}\label{eq:indcoupl}\frac{I^\prime}{I}\sim\frac{\omega L}{c^2R}\end{equation}
Here $\omega$ is the frequency, $c$ is the speed of light, and $R$ is the resistance of the bond. Assuming as a reference, values $L\sim 1$ cm, $R\sim 1$ Ohm,  and $\omega \sim 1$ GHz, the latter ratio is of the order of 1. However, lower frequencies and especially higher resistances (typically above 10$^3$ Ohm with PCM and RRAM applications) make that ratio small and acceptable.

(4) The characteristic $RC$ times related to the writing and reading processes can be estimated as $\sim 10$ ps for $R\sim 1$ Ohm and $\sim C\sim 10$ pF (corresponding to a 1 cm sample with the same macroscopic resistivity as that of 1 MOhm RRAM resistor with 10 nm linear dimension), rather competitive against the background of modern technology.

(5) The property of plasticity takes place when the local electric field exceeds its material dependent threshold value for resistance switching as discussed next.

\subsection{Large PWPs}\label{sec:LPWP}
The concept of infinite percolation cluster survives if the electrode sizes $l$ and interelectrode distances $L_{ij}$ are much larger than $L_c$, in which case (we call it `large PWP') resistances $R_{ij}$ are determined, in the ohmic regime, by electrode geometry,
\begin{equation}R_{ij}=(\sigma l)^{-1}f_{ij}(l/L_{ij})\label{eq:RL}\end{equation}
similar to the case of steady currents between finite size electrodes in massive conductors, such as grounding electrodes in a soil. Here $f_{ij}$ is a dimensionless function whose shape depends on the electrode locations through the confinement of electric currents by the sample boundaries. The macroscopic conductivity $\sigma$ in the equation for $R_{ij}$ is taken in the limit of infinitely large percolation systems where it is uniquely defined.

\begin{figure}[hb]
\includegraphics[width=0.43\textwidth]{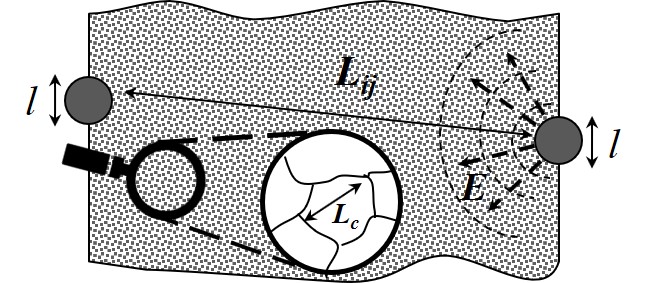}
\caption{A large PWP geometry with two hemisphere electrodes and a fragment of percolation cluster resolved in the magnifying glass. Shown in dash are the equipotential (arcs) and electric field (arrows) lines in the proximity of the right electrode illustrating the geometrical effect of electric field spreading with distance from the electrode. \label{Fig:LargePWP}}
\end{figure}
While $f_{ij}$ can be numerically modeled with for particular electrode configurations, some general statements can be made based on the available examples and a simplified model. \cite{hong2019} The latter presents an electrode as a metal hemisphere immersed into a macroscopically uniform medium formed by the percolation cluster over large scales as illustrated in Fig. \ref{Fig:LargePWP}. A rough estimate is given by  \begin{equation}f_{ij}\approx 1+O(l/L_{ij})+O(l/L)\label{eq:fij}\end{equation}
to the accuracy of a numerical multiplier that depends on a particular electrode geometry. Here $O(x)$ means `of the order of $x$' and it is assumed that $l/L\ll 1$, $l/L_{ij}\ll 1$.

It follows from Eqs. (\ref{eq:RL}) and (\ref{eq:fij}) that resistances $R_{ij}$ do not depend strongly on the interelectrode distances in the ohmic regime. Yet, the differences in the interelectrode resistances  will exist due to statistical fluctuations in their connecting bonds. The bond of length $L_{ij}$ in a large PWP will contain $L_{ij}/L_c$ quasi-independent cells of the percolation cluster. Each cell has on average resistance $R_c$ making the average bond resistance $\langle R_{ij}\rangle =R_cL_{ij}/L_c$.  Taking into account that resistances of individual cells exhibit random variation of the order of $R_c$, one thus arrives at the estimate for the characteristic relative variations of resistances,
\begin{equation}\label{eq:deltaR}\delta R/\langle R_{ij}\rangle\sim  \sqrt{L_c/L_{ij}}.\end{equation}
While relatively small, the fluctuations $\delta R/\langle R_{ij}\rangle$ are still significant enough to experimentally discriminate between different interelectrode resistances. For example, $\delta R/\langle R_{ij}\rangle\sim 0.1$ assuming $L_c\sim 10$ nm and $L_{ij}\sim 1 $ $\mu$m.

Unlike the standard percolation between two flat electrodes, in large PWP the electric field systematically decays with distance $r\gtrsim l$ from a small electrode due to the current spreading (again, similar to the case of grounding electrodes) as illustrated in Fig. \ref{Fig:LargePWP}. That geometrical effect will significantly alter the nature of nonohmicity making it most significant in the proximity of $r\sim l$ around the electrode. One can show that due to that field suppression, the steady state non-ohmic conduction will be limited to
\begin{equation}\label{eq:LPWP}r\sim l\sqrt{\xi_{\rm max}qaU/LkT}\end{equation}
where $U$ is the total voltage applied to a sample of length $L$. The pulse regime limitation will be described by a similar formula with the  substitution $\xi_{\rm max}\rightarrow \xi _t$.

\subsection{Small PWPs}\label{sec:SPWP}

The concept of infinite cluster fails when $l$ and/or $L_{ij}$ are smaller than $L_c$ (`small PWP'), in which case one has to consider multiple conductive paths unrelated to the infinite cluster. Both the cases of large and small PWP are possible across a broad variety of percolation systems. For example, the current $L\sim 10$ nm-node technology belongs in small PWP with \cite{patmiou2019} $a\gtrsim 0.3$ nm and $V_0/T\lesssim 100$, i. e. $L_c\sim 30$ nm. With the latter parameter values, increasing $L$ to microns and beyond will result in large PWP networks.

The resistances of conductive pathways in small PWP exhibit significant variability. Next, we estimate their statistics. Based on Eq. (\ref{eq:nonohm}) a chain resistance is estimated as $R=R_{\rm max}\exp(-\delta V/kT)$ where $\delta V=V_{\rm max}-v_{\rm max}$ and $v_{\rm max}$ is the maximum barrier in that chain, $R_{\rm max}=R_0\exp(V_{\rm max}/kT)$, and $V_{\rm max}$ is the maximum barrier in the entire system, $V_{\rm max}\geq v_{\rm max}$. Assuming uniformly distributed barriers, the average number of resistors with the barriers above a given $v_{\rm max}$ in a $n$-resistor chain is $n_v=n(V_{\rm max}-v_{\rm max})/(V_{\rm max}-V_{\rm min})$, where $V_{\rm min}$ is the minimum barrier. The Poisson probability of a chain having no barriers greater than $v_{\rm max}$ is $P_V(n)=\exp(-n_v)$, and the probability of finding a chain with a maximum barrier in the interval $kT$ around $v_{\rm max}$ is
\begin{equation}\label{eq:pvmax}P(v_{\rm max})=P_V(n)kT/(V_{\rm max}-V_{\rm min}).\end{equation}

Multiplying $P(v_{\rm max})$ by the probability $P_n(L)\sim (L/an)\exp(-L^2/na^2)$ of an n-resistor chain connecting points distance $L$ from each other, we obtain the probability density of n-chain with a given barrier $v_{\rm max}$ (to the accuracy of a numerical multiplier in the exponent). Integrating that product over $n$ by steepest descent and expressing $v_{\rm max}$ through $R$ yields the probabilistic distribution density,
\begin{equation}P(R)\propto \frac{1}{R}\exp\left(-\frac{2L}{a}\sqrt{\frac{kT}{V_{\rm max}-V_{\rm min}}\ln\frac{R_{\rm max}}{R}}\right).\end{equation}
It follows that resistance spectrum is a gradual function with a certain characteristic width $\Delta R$. As estimated separately for the cases of small and large $L$, width $\Delta R$ can be approximated for the entire range of $L$ by the following equation:
\begin{equation}\Delta R\approx R_{\rm max}\left[1+\left(\frac{L}{a}\right)^2\frac{kT}{V_{\rm max}-V_{\rm min}}\right]^{-1}.\end{equation}
For all practical values, it encompasses multiple orders of magnitude.

Because of the dispersion in the values $v_{\rm max}$ between different pathways, the nonohmicity exponents will vary from one $R_{ij}$ to another. More specifically, instead of $\xi _{\rm max}$ in Eq. (\ref{eq:nonohm}) the value $v_{\rm max}/kT$ should be used with the probability distribution of Eq. (\ref{eq:pvmax}). That additionally broadens the distribution of path resistances in the nonohmic regime; we omit here the obvious formal description of that effect.

\subsection{Plasticity by switching}\label{sec:plast}

A unique non-ohmicity feature of PWP is its nonvolatile nature rendered by the underlying material (say, of PCM or RRAM type). Each microscopic element of a conductive path can exist in either  high or low-resistive state whose respective resistances, $R_{>}$ and $R_{<}$, are markedly different. $R_>$ resistances are random, all exceeding $R_<$. The applied bias concentrated on the strongest of $R_{>}$ resistors (in the manner of Fig. \ref{Fig:PWP2}) will change them to $R_{<}$ {\it by switching}, i. e. by long-lived structural transformation not adaptable to subsequent voltage variations. In the steady state bias regime, the next strongest resistor will be stressed with practically {\it the same voltage} as opposed to the above discussed case of volatile non-ohmicity in the standard percolation clusters. In the first approximation, an originally resistive  percolation bond will transform into its conductive state by $n$ discrete steps where $n$ is the number of its microscopic resistances, similar to a falling row of dominoes arranged in the order of descending $\xi$'s.

The latter behavior can be more complex in large PWPs due to the geometrical field distortion illustrated in Fig. \ref{Fig:LargePWP}. There, the field will eliminate large resistances in the region of characteristic length given by Eq. (\ref{eq:LPWP}) which then becomes a sort of low resistive protrusion into the bulk material. Such a protrusion will concentrate the electric field similar to the lightning rod effect. As a result, switching will take place in the next high field domain growing that protrusion further, etc., until it reaches the opposite electrode.  While the kinetics of such a process can be readily described, we will omit it here.

For switching to occur, the local field on a microscopic resistor must exceed a certain critical value \cite{legallo2020,karpov2008} $E_c$, typically on the order of $10^5-10^6$ V/cm, and $E_c$ decreases logarithmically with the electric pulse (spike) length. \cite{karpov2008,krebs2009,bernard2010,sharma2015,you2017} That temporal dependence opens a venue to the spike timing dependent plasticity (STDP), \cite{markram2011} which is another important property of neural networks. Because almost the entire voltage drops across a microscopic resistance of small linear size $a$, the microscopic field $E$ is significantly stronger than the apparent macroscopic field and was shown \cite{karpov2020} to approach $\sim 20$ MV/cm, well above the values of $E_c$ sufficient for threshold switching.

\subsection{Plasticity in the pulse regime}\label{sec:pulse}
Pulse excitation regime brings in additional physics. \cite{karpov2020} The random elements of a percolation bond will behave as resistors when their relaxation times $\tau $ are shorter than the pulse duration $t$; however the elements with $\tau\gg t$ will act as capacitors causing no significant voltage drop in response to short pulses. The boundary between these fast and slow elements is defined by $\tau = t$, i. e. $\xi =\xi _t\equiv \ln(t/\tau _0)$ given that $\tau =\tau _0\exp(\xi)$.

\begin{figure}[hb]
\includegraphics[width=0.37\textwidth]{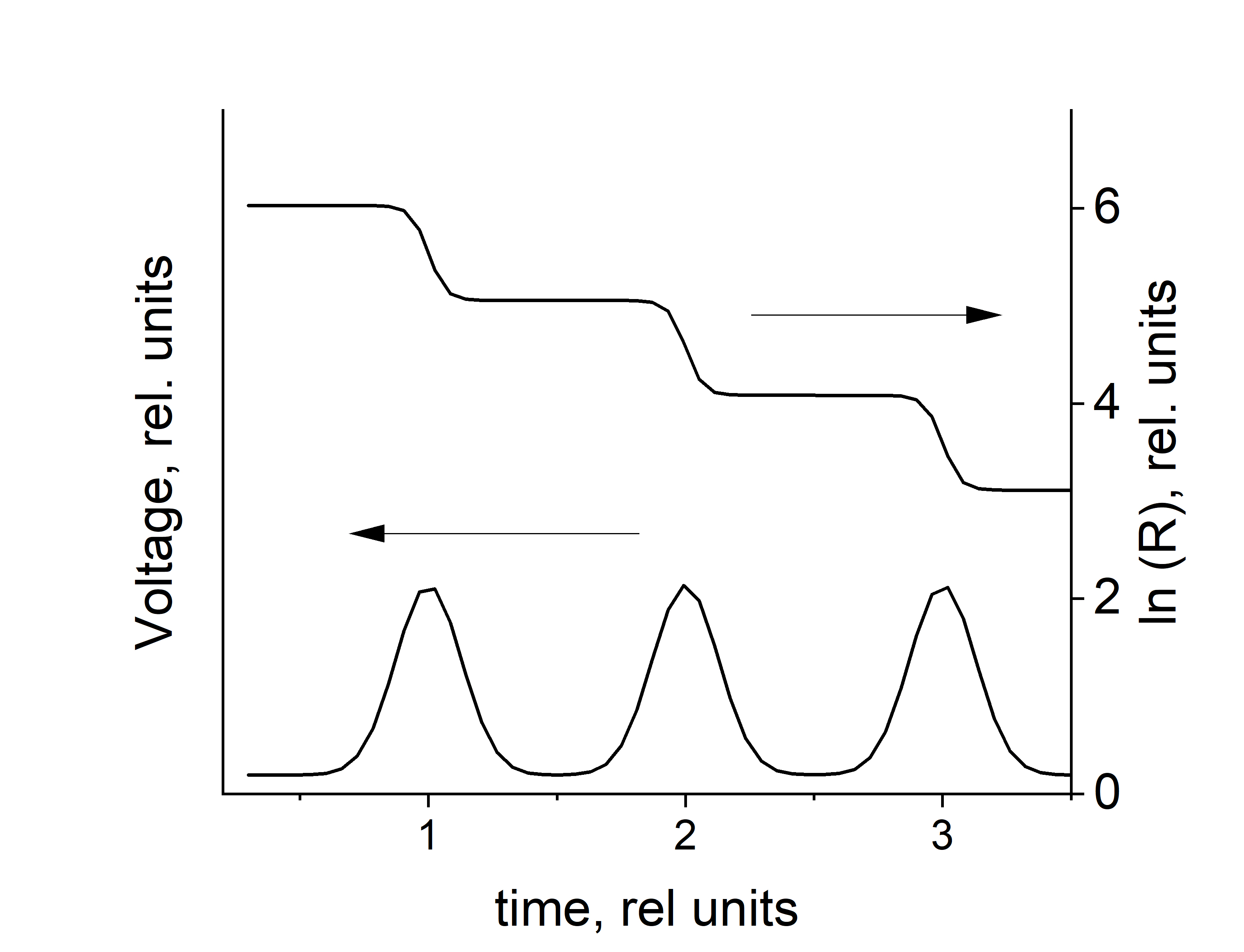}
\caption{A series of stepwise drops in a PWP bond resistance in response to a train of pulses. \label{Fig:PWPsteps}}
\end{figure}

The resistance $R_t=R_0\exp(\xi _t)$ will be the highest of all bond elements resistances and the voltage pulse will concentrate on it for its duration. Should the corresponding local field exceed $E_c$, the resistance $R_t$ will switch to $R_<$ with other resistors intact, which leads to the interelectrode path resistance decreasing by roughly a factor of $[e]$ (base of natural logarithms) as illustrated in Fig. \ref{Fig:PWPsteps}. It is straightforward to show that the number of such stepwise changes in a bond resistance is estimated as
\begin{equation}\label{eq:M}\sqrt{qV/kT}\sim 10,\end{equation} which property can, in principle, be used to create multi-level memory operated by trains of pulses.

Corresponding to Eq. (\ref{eq:M}), the characteristic voltage per microscopic resistor is given by,\cite{karpov2020}
\begin{equation}\label{eq:U1}U_1\approx U_1=U_L\sqrt{\frac{kT}{qU_L}},\end{equation}
where $U_L$ is the macroscopic voltage across the bond of length $L$. According to the numerical estimates in Sec. \ref{sec:num} below, $U_1$ can create the microscopic field of strength exceeding the switching value.

The concept of slow resistors acting in a manner of capacitors has been proposed and verified earlier. \cite{dyre2000,abje2016} It may be appropriate to additionally explain here that capacitors do not accommodate significant voltages when in series with resistors because they conduct the displacement currents, $j_D=(\varepsilon /4\pi )(d{\cal E}/dt)$, where $\varepsilon$ is the dielectric permittivity. The overlay between the capacitor and resistor regimes takes place when $j_D= j=\sigma {\cal E}$ where $j$ is the real (charge transport) current, ${\cal E}$ is the electric field strength, and $\sigma $ is the conductivity. The displacement current through a capacitor is due to the rate of field change, unrelated to voltage, rather than the field itself proportional to voltage in a resistor.

The ratio of displacement vs. real current can be presented as $j_D/j=(\varepsilon /4\pi\sigma )(d\ln {\cal E}/dt)$. The expression in the first parenthesis represents the Maxwell's relaxation time $\tau$, while the reciprocal of the second parenthesis gives the pulse duration. That takes us again to the criterion $\tau \gg t$ for the element of a percolation cluster operating in the capacitive mode.

Relating this understanding with microscopic models, we note that the displacement currents are due to charging/discharging processes in, say, capacitor electrodes, or in certain defect configurations responsible for electric potential distributions in percolation clusters.

\subsection{Reverse plasticity}\label{sec:reverse}
The above description does not address the important question of reversibility of switched structures back to high resistive states. The feasibility of such reverse process follows from the known practices of RRAM and PCM operations. More specifically, two conceivable answers include thermal annealing towards the original high resistive state by Joule heat generated by relatively low currents over considerable times.  Another possibility is based on PWPs built of materials with a degree of ferroelectricity allowing reversibility in response to electric polarity changes. \cite{karpov2017}

\section{Examples of functionality}\label{sec:examp}

{\it Multivalued memory.} When high enough voltage is applied between a pair of PWP electrodes, their connecting path will change its resistance depending on the pulse duration as explained in Sec. \ref{sec:plast}. Given for example $N=3$ electrodes on each of the faces of 1x1x1 cm$^3$ cube, the number of perceptive pathways $M=N!\sim 10^{16}$  cm$^{-3}$ is {\it theoretically} higher than that of human cortex resulting in a higher memory capacity than the current 3D crossbar architecture. It will be further enhanced with the functionality of multiple records per one electrode pair. We should admit however that various unaccounted factors could interfere and the claim of that superior memory capacity remains to be validated experimentally.

{\it Generation of random numbers.} As shown in Sec. \ref{sec:PWPST}, not-too-spatially-close pairs of electrodes have random resistances uncorrelated with any desired accuracy when their spacial separation increases. They form a multitude of uncorrelated random numbers.

{\it Matrix-vector multiplication.} The measurement based operation of matrix-vector multiplication follows from Fig. \ref{Fig:PWPconcept}. Suppose that $A_i$ is the desired product of the vector $J_j$ and the matrix $F_{ij}$. We rescale $J_j$ with a certain multiplier ($z_1$) to a convenient interval of electrode voltages ${\cal E}_j$. Secondly, using a proper multiplier ($z_2$), we rescale $F_{ij}$ so that all its elements fall in the interval of PWP system conductances,  $\delta G=\delta R/\langle R_{ij}^2\rangle$ with $\delta R$ from Eq. (\ref{eq:deltaR}).  The desired product becomes $A_i=z_1z_2I_i$ where $I_i=\sum _jG_{ij}{\cal E}_j$ is the current through the $i$th electrode in Fig. \ref{Fig:PWPconcept}. Because the conductance matrix $G_{ij}=R_{ij}^{-1}$ contains exponentially large number ($M\gg 1$) of elements covering interval $\delta G$, any desired value of $G_{ij}$ can be located at least approximately  among the measured conductances with fairly good accuracy {\it without any additional actions}. After that, applying voltage ${\cal E}_j$ to the electrode $j$ produces a measurable current $G_{ij}{\cal E}_j$ through electrode $i$, which can be stored e. g. as a partial charge on a certain capacitor $C_i$. Measuring the total of all such partial contributions supplied by electrode $i$ in response to various ${\cal E}_j$ will give the component of sought vector $I_i$. That procedure can be further improved by choosing different approximate $G_{ij}$s and using linear regression for all the chosen values.

{\it Brain-like associative learning} commonly illustrated with Pavlov's dog salivation experiments (see e. g. Ref. \onlinecite{kuzum2013}) is readily implemented utilizing shared portions between bonds of a PWP cluster, such as bonds (1,5) and (1,7) in Fig. \ref{Fig:PWPconcept}. Identifying the `sight of food' and `sound' stimuli with signals on the electrodes 5 and 7, predicts that properly and simultaneously triggering both will switch their corresponding pathways (1,5) and (1,7) to a low resistive state making both salivation triggering (through the output on electrode 1).  In general, conductive pathways connecting various pairs of the electrodes and sharing the same portion of a PWP cluster will be mutually affected by a single bias-induced change. That demonstrates a single-trial learning model for storage and retrieval of information resembling that of the cortex of the mammalian brain.

{\it Other functionalities} based on PCM and RRAM structures for neuromorphic computing appear all attainable with PWP systems. We note that even without utilizing their plasticity, PWPs can serve as high capacity tunable nonlinear reservoirs for reservoir computers. For example, introducing nonohmic (yet volatile) changes in the resistances of pathways (1,5), (11,5), and (5,7) will produce measurable changes in the resistances of pathways (11,3), (1,7), and (11,7), and the latter  will depend on temporal order in which the former changes were introduced. Utilizing the system plasticity will significantly add to that functionality. In fact, our proposed PWPs represent exponentially more powerful reservoir computing systems compared to the one built using a limited number of memristors. \cite{tanaka2017}

\section{PWP metrics and implementations}\label{sec:met}
	
We briefly mention several metrics of the proposed PWP devices following the nomenclature used for other neuromorphic systems. \cite{kuzum2013}  (1) {\it Dimensions and architecture.} The above estimate of a superior information density may be reduced to account for larger physical dimension of a single microscopic resistor. However, even assuming a microscopic resistor of PWP $a$ in the range of tens of microns yields the density $\sim 10^9$ cm$^{-3}$. (2) {\it Energy consumption}. Assuming a PWP strucure made of the same materials as the existing PCM and RRAM, we expect its energy efficiency to be superior because of the lack of interconnects requiring costly energy. (3) {\it Operating speed/programming time}. Generally, PWP devices RC times are greater than those of nano RRAM and PCM.  Like other brain-inspired systems, their computational efficiency will be achieved through the high degree of parallelism. We recall in this connection that the combinatorial huge number of memory units $N!$ is exponentially higher that the number of electrodes $N$. (4) {\it Multi-level states}. Assuming $a\sim$ 10 nm microscopic resistor and 1 cm device, each bond in a PWP cluster will contain hundreds of micro-resistors; hence, hundreds of multi-level states per typical bond, at the level allowing robust analog operations. (6) {\it Retention and endurance}. PWP systems can be superior to the existing PCM/RRAM based devices because of the lack of multiple interconnects triggering degradation.

\section{Similarity with biological systems}\label{sec:bio}

The PWP architecture and functionality have similarities with that of biological neural networks. We recall that the latter consists of individual neurons integrating input signals and firing pulses upon exceeding a certain threshold of integration. The neurons communicate with network by means of synapses that inject ions through their membrane ionic channels then electrically altering adjacent neurons. The neuro-synaptic entities are connected with each other through axons providing pathways for electric pulses. A significant degree of randomness and stochasticity is introduced by ion channels whose concentrations and characteristics vary even between nominally similar biological membranes. \cite{elkins2013,liu2001,golowasch2014,srikanth2015}

The property of firing pulses upon accumulating enough electric charge is found in PWP's slow elements ($\tau >t$) that operate as capacitors. Such elements turn into resistors when, through their multiple connections in PWP, they acquire voltages sufficient to make real currents larger than the displacement ones. \cite{karpov2020} Such accumulation becomes possible when the signals arrive within a certain time interval thus resembling STDP in neural networks. The inherent randomness of PWP correlates with that in biological neural networks.

Note that the neuron analogy described here pertains to the system functionality, but not the structure. In PWP, slow elements can be associated with any of the system elements, depending on the pulse duration. Therefore, the PWP neuron-like elements are in a sense distributed throughout the system and the same element can play the role of a synapse, or axon, or neuron depending on the excitation conditions. The concept of functional similarities between PWP and biological neural networks as outlined here remains in its infancy calling upon further analyses.

\section{Numerical estimates of PWP parameters}\label{sec:num}
Assuming $a\sim 1$ nm and $V_{\rm max}/kT\sim 100$ yields $L_c\sim aV_{\rm max}/kT\sim 100$ nm. Empirically, the field reliably leading to switching is $\lesssim 1$ MV/cm. On the other hand, it is estimated by our theory as  $U_1/a$, with $U_1$ from Eq. (\ref{eq:U1}), which yields $U_L\lesssim 0.1$ V. The latter corresponds to a fairly attainable electric potential drop of $\lesssim 10$ kV across 1 cm thick samples.

We conclude that multivalued memory by switching in PWP corresponds to rather moderate voltages at least for large PWP systems. Because the earlier estimated $\sim 10$ records corresponds to the bond of correlation radius $L_c\sim 100$ nm, the number of records per 1 cm thick sample is estimated as $\sim 10^6$. The question of their temporal overlaps remains to be addressed based on the known statistical properties of percolation clusters.

For small PWP, using the same parameters and sample size $L\lesssim L_c$ one obtains $\sim 10L/L_c $ possible transitions, which however will be all distinctly different. Note that for 10 nm samples that estimate predicts one switching event, which on average is what is observed in the current RRAM and PCM devices.

\section{The role of randomness}\label{sec:disorder}
A characteristic feature of our proposed PWPs is that they are generically random and possess exponentially broad spectra of parameters. As such, they are suggestive of  probabilistic algorithms that employ a degree of randomness as part of its logic. \cite{yao1997} We are not aware of any general concept substantiating the advantages of randomness for computations, although the challenge of it is quite appreciated by now and became a hot topic. \cite{randomness,mitzenmacher2017}

The uniqueness of our approach is that it proposes a generically `random hardware' with which even a logically deterministic algorithm becomes randomized. The general power of such approach remains to be tested yet, although some examples do point at its potential: superiority of randomly-wired networks, \cite{hie2019,zopf2018}  and intentionally randomized reservoir computers. \cite{tanaka2019} Biological neural networks provide of course an ultimate example of neuromorphic computing leveraging randomness (we are not built of well controlled silicon nano-chips).

From that perspective, our approach proposes an interesting and practically implementable machinery for testing the potential of `randomized neuromorphic hardware'. That goal can be achieved both through computer modeling of PWP that benefits from the earlier developed percolation modeling algorithms, and through direct material implementations, such as e.g chalcogenide films with multiple electrodes; both directions are being attempted by our group.

\section{Conclusions}\label{sec:concl}

We have introduced the percolation with plasticity (PWP) systems that, while being essentially random, exhibit various neuromorphic functionalities and similarities to the cortex of mammalian brain. These systems demonstrate rich physics, understanding of which remains in its infancy.

The outstanding challenges in theory go far beyond the standard percolation paradigm including statistical aspects, microscopic phase transformations, heat transfer, AC propagation and signal cross-coupling in random systems with multiple co-existing interfaces. Both analytical research and numerical modeling will help to better understand PWP operations and functions.  Multiple material bases can be used to fabricate PWP devices, ranging from the known PCM and RRAM materials to the nanocomposites which beg to be further explored for memory and other neuromorphic applications. The above noted functional similarities with biological neural networks suggests the need for further investigations.

To avoid any misunderstanding we should emphasize that the present work is limited to the basic theory of PWP systems. It only tangentially addresses the related neuromorphic applications that can be tried at both the levels of software implementations and real material devices. Obvious experimental projects potentially triggered by this work include pulse regime nonohmicity in disordered systems, statistical properties of percolation clusters with multiple electrodes, and reservoir computing using with percolation systems.

\section*{Acknowledgements}
We are grateful to A. V. Subashiev and D. Shvydka for useful discussions.

\end{document}